\documentclass[manuscript,screen]{acmart}

\usepackage[nolist,nohyperlinks]{acronym}
\usepackage{amsmath}
\AtBeginDocument{%
  \providecommand\BibTeX{{%
    \normalfont B\kern-0.5em{\scshape i\kern-0.25em b}\kern-0.8em\TeX}}}

\copyrightyear{2021}
\acmYear{2021}
\setcopyright{acmlicensed}
\acmConference[RecSys '21]{Fifteenth ACM Conference on Recommender Systems}{September 27-October 1, 2021}{Amsterdam, Netherlands}
\acmBooktitle{Fifteenth ACM Conference on Recommender Systems (RecSys '21), September 27-October 1, 2021, Amsterdam, Netherlands}
\acmPrice{15.00}
\acmDOI{10.1145/3460231.3474230}
\acmISBN{978-1-4503-8458-2/21/09}

\usepackage{multirow}
\begin{document}
\title{Hierarchical Latent Relation Modeling for~Collaborative~Metric~Learning}

\author{Viet-Anh Tran}
\authornote{Contact author: \href{research@deezer.com}{research@deezer.com}}
\affiliation{
  \institution{Deezer Research}
    \city{}
  \country{France}
}

\author{Guillaume Salha-Galvan}
\affiliation{
  \institution{Deezer Research \& LIX, \'{E}cole Polytechnique}
  \city{}
  \country{France}
}

\author{Romain Hennequin}
\affiliation{
  \institution{Deezer Research}
    \city{}
  \country{France}
}

\author{Manuel Moussallam}
\affiliation{
  \institution{Deezer Research}
    \city{}
  \country{France}
}

\renewcommand{\shortauthors}{V.A. Tran et al.}


\begin{acronym}
    \acro{CML}{Collaborative Metric Learning}
    \acro{MF}{Matrix Factorization}
    \acro{MAP}{Mean Average Precision}
    \acro{MRR}{Mean Reciprocal Rank}
    \acro{DCG}{Discounted Cumulative Gain}
    \acro{NDCG}{Normalized Discounted Cumulative Gain}
    \acro{CF}{Collaborative Filtering}
    \acro{LRML}{Latent Relation Metric Learning}
    \acro{HLR}{Hierarchical Latent Relation}
    \acro{JUPITER}{Joint User Preferences and Item Relations}
    \acro{TransCF}{Collaborative Translational Metric Learning}
    \acro{AdaCML}{Adaptive Collaborative Metric Learning}
    \acro{FISM}{Factored Item Similarity Model}
    \acro{KG}{Knowledge Graph}
\end{acronym}

\begin{abstract}
  \ac{CML} recently emerged as a powerful paradigm for recommendation based on implicit feedback collaborative filtering. However, standard \ac{CML} methods learn fixed user and item representations, which fails to capture the complex interests of users. Existing extensions of \ac{CML} also either ignore the heterogeneity of \textit{user-item} relations, i.e. that a user can simultaneously like very different items, or the latent \textit{item-item} relations, i.e. that a user’s preference for an item depends, not only on its intrinsic characteristics, but also on items they previously interacted with. In this paper, we present a hierarchical CML model that jointly captures latent \textit{user-item} and \textit{item-item} relations from implicit data. Our approach is inspired by translation mechanisms from knowledge graph embedding and leverages memory-based attention networks. We empirically show the relevance of this joint relational modeling, by outperforming existing \ac{CML} models on recommendation tasks on several real-world datasets. Our experiments also emphasize the limits of current \ac{CML} relational models on very sparse datasets.
\end{abstract}


%% \ccsdesc{Computer systems organization~Robotics}
%% \ccsdesc[100]{Networks~Network reliability}


\maketitle

%%%%%%%%%%%%%%%%%%%%%%%%%%%%%%
%       INTRODUCTION
%%%%%%%%%%%%%%%%%%%%%%%%%%%%%%
\section{Introduction}
\label{s1}

Nowadays, it is common for users to face the situation of \textit{“too much content, too little time”} when listening to music or watching videos on streaming services with large catalogs \cite{schedl18:jmir,briand21:kdd}, or when purchasing products on e-commerce websites~\cite{schafer01:dmkd}. These online services heavily rely on recommender systems to address this information overload, by identifying the most relevant content for each user. They are known to be key components to attract and engage users, and are also central to all enjoyable passive experience relying on generated content \cite{bobadilla13:kbs,schedl18:jmir,zhang2019deep}.

%Users often face the \textit{“too much content, too little time”} problem when listening to music or watching videos on streaming services with large catalogs \cite{schedl18:jmir}, or when purchasing products on e-commerce websites \cite{schafer01:dmkd}. These online services heavily rely on recommender systems to address this information overload, by identifying the most relevant content for each user. They are known to be key components to attract new users, and improve their engagement on the service \cite{bobadilla13:kbs}.

A prevalent paradigm to recommend personalized content on these services is \ac{CF}. Operating on \textit{implicit feedback data}, such as streams or skips for music streaming, \ac{CF} methods assume that users who had similar interests in the past will tend to share similar interests in the future \cite{koren2009matrix,bobadilla13:kbs}. In particular, \ac{MF} remains one of the most popular \ac{CF} baselines \cite{koren2009matrix,hu08:icdm, rendle09:auai}. It consists in learning low-dimensional vector representations of users and items from the factorization of an interaction matrix summarizing feedback data, and subsequently modeling the future user–item similarities from inner products between these vectors.

However, \ac{MF} has recently been competed by \textit{metric learning} methods \cite{khoshneshin10:ijcai, chen12:sigkdd, feng15:ijcai, hsieh17:www, zhang18:corr, paudel18:wsdm, zhou19:ijcai, tran19:sigir} such as Collaborative Metric Learning (\ac{CML}) \cite{hsieh17:www}, where similarities are instead measured by Euclidean or Hyperbolic distances \cite{tran20:wsdm}. Specifically, \ac{CML} adopts a \textit{triplet loss} strategy \cite{weinberger:jmlr09, song17:aaai} to learn vector representations ensuring, as explained in Section \ref{s2}, that distances between users and their \textit{negative items} (which they did not interact with) are larger than distances between users and their \textit{positive items} (which they interacted with).
%by a fixed margin of $m > 0$. That is, $d(p_u, q_{v}) + m \leq d(p_u, q_{v^-})$ where $d$ is the positive distance function between two entities and $p_u$, $q_{v}$, $q_{v^-}$ are vectors of user, positive item and negative item respectively. 
Relying on the assumption that users should be closer to items they like, this approach implicitly learns user-user and item-item similarities by satisfying~the~triangle~inequality~\cite{hsieh17:www}.

\ac{CML} nonetheless suffers from an inherent limitation: each user is represented \textit{by a single vector}, which does not fit the many-to-many nature of real-world recommendation problems \cite{tay18:www, park18:icdm, zhang19:dsaa, zhou19:ijcai}. In particular:
\begin{itemize}
    \item Standard \ac{CML} fails to capture the \textit{heterogeneity in user-item relations} \cite{tay18:www,park18:icdm, zhou19:ijcai}. Indeed, due to the triangular inequality, a user cannot be simultaneously represented as close to two items that are themselves distant. This is undesirable as, in reality, a same user could simultaneously like very different movies or music genres.
    %\item Such a restriction also hinders modeling the \textit{intensity of user-item relations} in implicit feedback \cite{park18:icdm}, i.e. that a user’s similar feedback on two items (e.g. a stream) might not always express an equal preference for~these~items.
    \item Moreover, \ac{CML} does not model the fact that a user’s preference for an item depends, not only on its intrinsic characteristics, but also on items they previously interacted with \cite{zhang19:dsaa}, which we refer to as \textit{item-item relations}.
\end{itemize}

While some extensions of \ac{CML}, described thereafter in Section \ref{s2}, aimed at modeling user-item relations \cite{tay18:www,park18:icdm, zhou19:ijcai}, they did not consider the  item-item ones. On the contrary, recent efforts on item-item relations-aware~\ac{CML}~\cite{zhang19:dsaa} ignored user-item relations.
In this paper, we argue that \textit{jointly} modeling both user-item and item-item relations can actually help revealing the underlying spectrum of user preferences, and thus lead to better performances in recommendation. More specifically, our work builds upon the assumption, presented in Section \ref{s3}, that there exists \textit{a hierarchical structure} in different relation types, and that \textit{user-item relations are built on top of item-item relations}.

This assumption is materialized in the form of a neural architecture, named \ac{HLR}, that leverages the recent advances in attention mechanisms, augmented memory networks as well as metric learning to explicitly learn adaptive user-item and item-item relations lurked in implicit feedback data. We provide comprehensive experiments on four real-world datasets, demonstrating the relevance of our proposed approach and its empirical effectiveness, in terms of recommendation accuracy, ranking quality, and popularity~bias~\cite{schedl18:jmir}.
Along with this paper, we also publicly release our source code on GitHub to ensure the reproducibility of our results.

This paper is organized as follows. In Section \ref{s2}, we formulate our recommendation problem more precisely, and recall key concepts on \ac{CML} and its existing extensions. In Section \ref{s3}, we present our hierarchical system to jointly model user-item and item-item relations. We report and discuss our experiments in Section \ref{s4}, and we~conclude~in~Section~\ref{s5}.
%%%%%%%%%%%%%%%%%%%%%%%%%%%%%%
%         CONTENT
%%%%%%%%%%%%%%%%%%%%%%%%%%%%%%
\section{Background}
\label{s2}
\subsection{Problem Formulation}
\label{s21}
In this paper, we consider a top-$K$ recommendation problem on an online service, with a set of users $\mathcal{U}$ and a set of recommendable items $\mathcal{V}$. We observe some positive implicit and binary user feedback (e.g., clicks, streams or view history logs) on items, gathered in a set $\mathcal{S} = \{(u, v) | u \in \mathcal{U}, v \in \mathcal{V}\}$. If we interpret $\mathcal{U}$ and $\mathcal{V}$ as two disjoint and independent sets of vertices in a bipartite user-item graph $\mathcal{G}$, then $\mathcal{S}$ represents the ensemble of edges that connect a vertex in $\mathcal{U}$ to one in $\mathcal{V}$. Missing interactions do not necessarily imply a negative feedback; in most cases, the user is simply unaware of the items' existence. We use $\mathcal{N}_u \subseteq \mathcal{V}$ to denote the set of items that user $u$ previously interacted with. In such a setting, we aim at recommending $K>0$ items $v \in \mathcal{V} \setminus \mathcal{N}_u$ to each user $u$, for which they should positively interact with. The problem can thus be framed as a missing link prediction task in the bipartite graph~$\mathcal{G}$.

\subsection{Collaborative Metric Learning}
\label{s22}

To address this top-$K$ recommendation problem, \ac{CML} \cite{hsieh17:www} learns a \textit{metric space} encoding $\mathcal{S}$ and representing both users and items as low-dimensional vectors. In such a space, user-item similarities are measured with the Euclidean distance $ d(u, v) = ||p_u - q_v||^{2}_2$, where $||\cdot||_2$ denotes the L2-norm, and where $p_u$ and $q_v$ are the respective representation vectors of user $u$ and item $v$ in the metric space. At its heart, \ac{CML} aims at pulling closer matching user-item pairs in $\mathcal{S}$, and at pushing away non-matching pairs, up to a certain margin $m > 0$. Formally, let us denote by $\mathcal{T}$ the ensemble of \textit{triplets} $(u, v, v^-)$ such that the user $u$ interacted with item $v$ and did not interact with item $v^-$, i.e. $(u,v) \in \mathcal{S}$ and $(u,v^-) \notin \mathcal{S}$. Let us denote by $p_u$ , $q_v$ and $q_{v^-}$ their respective vectors. Then, \ac{CML} essentially aims at solving:
\begin{equation}
    \min_{||p_u||_2 \leq 1, ||q_v||_2 \leq 1, ||q_{v^-}||_2 \leq 1} \sum_{(u, v, v^-) \in \mathcal{T}} \Big[||p_u - q_v||^{2}_2 - ||p_u - q_{v^-}||^{2}_2 + m\Big]_+
    \label{triplet_loss}
\end{equation}
where $m>0$ is a fixed margin hyperparameter, and $[x]_+ = max(0, x)$ denotes the standard hinge loss. Finally, the top-$K$ items recommended to the user $u$ will correspond to its $K$ closest ones in the resulting space.

\subsection{Modeling User-Item Relations}
\label{s23}

\subsubsection{On the Limits of Single Vector Representations}

Despite promising results against popular alternatives including \ac{MF} \cite{hu08:icdm, rendle09:auai, he17:www}, standard \ac{CML} suffers from the fact that each user is represented \textit{by a single vector} in the metric space. As mentioned in Section \ref{s1}, the \textit{heterogeneity in user preferences} can not be satisfactorily handled under such restriction, as the triangular inequality imposes that $p_u$ cannot be simultaneously close to two items that are distant. For instance, a user could simultaneously like metal and classical music, whereas songs from these two music genres are likely to be far from each other in a \ac{CML}-based metric space. Also, a single distance hardly captures the fact that a user might like a song or a movie for multiple reasons, such as its genre, its director, or its casting.

\subsubsection{User Translation in Vector Space}
To overcome these limitations, recent works \cite{tay18:www, park18:icdm} leveraged the concept of \textit{translation in vector space} to learn an adaptive \textit{relation vector} $r_{uv}$ for each user-item pair $(u,v)$. This concept takes inspiration from reasoning over semantics in a \ac{KG} embedding \cite{bordes13:nips}, where some entities and relations from a \ac{KG} are represented in a vector space. As an example, let us consider a \ac{KG} connecting geographical entities such as countries and cities.
Given a fact/an edge (e.g. "$u$ is the capital of $v$"), the semantic relation ("capital") would be captured by a translation vector $r$, in the sense that the embedding vectors of $u$ and $v$, say $p_u$ and $p_v$, would verify $p_u + r \approx p_v$.
Then, the plausibility of a semantic link between two new entities (e.g. "is $w$ the capital of $x$?") is measured from the distance between $p_w + r$ and $p_x$. TransE \cite{bordes13:nips} is a famous example of such a translational distance model.

\subsubsection{User-Item Relation Modeling for CML}
Transposing this concept to recommendation, extensions of \ac{CML} \cite{tay18:www,park18:icdm} proposed to learn metric spaces by minimizing $||p_u + r_{uv} - q_v||^2_2$ terms in equation (\ref{triplet_loss}), with a relation vector $r_{uv}$ for each user-item interaction. Notably, Tay et al. \cite{tay18:www} introduced \ac{LRML}, which learns $r_{uv}$ vectors by means of a weighted adaptive translation over an augmented memory via neural attention. This permits translating a user $u$'s vector ($p_u  \rightarrow p_u + r_{uv}$) based on the candidate item $v$, and therefore capturing heterogeneities in user preferences contrary to \ac{CML}. Zhou et al. \cite{zhou19:ijcai} then further extended this idea to multiple user-item~relation~types.

Another related model is \ac{TransCF} \cite{park18:icdm}, whose authors argued that the above approach might be prone to \textit{overfitting} due to the large number of parameters involved, and that it does not allow the collaborative information to be explicitly modeled as no parameters are shared among users and items. Authors of TransCF instead took inspiration from neighborhood–based \ac{CF} algorithms \cite{koren08:sigkdd, santosh13:sigkdd}, that assume that similar users display similar item preferences and that similar items are consumed by similar users. TransCF employs the neighborhood information of users and items to learn translation vectors $r_{uv}$. More precisely, each item is regarded as the average taste of users having interacted with that item. Likewise, each user’s taste can be represented by the average representation of items this user interacted with. Given the respective representations of an item and a user from the neighborhood perspective, the relation vector $r_{uv}$ is modeled by a simple element-wise vector product~of~these~two~vectors.

\subsection{Modeling Item-Item Relations}
\label{s24}

\subsubsection{On the Limits of Single Vector Representations} Besides user-item relations, single vector representations also fail to capture latent \textit{item-item relations}. This refers to situations where a user’s preference for an item depends, not only on its characteristics, but also on items they previously interacted with \cite{zhang19:dsaa}. For instance, some users could decide to interact with an item because of its similarity with several items in their favorites. The aforementioned \ac{LRML} and TransCF extensions of \ac{CML} also ignore this aspect. As they focus on a single user-item interaction at a time, they can not capture relations between a candidate item and items that a user previously interacted with.

\subsubsection{Item-Item Relation Modeling for CML}
Recent efforts from Zhang et al. \cite{zhang19:dsaa} aimed at learning adaptive relation vectors $r_{uv}$ that capture such latent item-item relations. Authors proposed \ac{AdaCML}, an extension of the popular FISM model \cite{santosh13:sigkdd}. As in FISM, users are represented by a linear combination of the items they interacted with in the past. However, while FISM considers that all items contribute equally when estimating the similarity between a user and a target item, \ac{AdaCML} learns the impact of each historical item via an attention mechanism. Attention weights depend on each target item and are derived from inner products of this item with historical items in users' favorite lists. The final relation vector $r_{uv}$ is then derived from a weighted sum of all historical items that the user interacted with. We emphasize that \ac{AdaCML} only concentrates on item-item relations, and does not model the heterogeneity lurked in each \textit{user-item} interaction.
\section{Hierarchical Latent Relation Modeling}
\label{s3}
In this section, we introduce our proposed Hierarchical Latent Relation (\ac{HLR}) model, and its extension \ac{HLR}++, to jointly learn user-item and item-item relations. Their overall architectures are summarized in Figure \ref{fig:hlr} and Figure \ref{fig:jupiter}.

\subsection{Motivations}

From Sections \ref{s23} and \ref{s24}, we conclude that existing extensions of \ac{CML} capturing user-item relations \cite{tay18:www,park18:icdm} did not consider the item-item ones. Simultaneously, extensions capturing item-item relations \cite{zhang19:dsaa} did not consider the user-item ones.
In this paper, we argue that user-item and item-item relations can actually complement and influence each other. This would imply that \textit{jointly} learning both relations could help reveal fine-grained user preferences, leading to better performances in recommendation.

In the remainder of this section, we leverage attention mechanisms and augmented memory networks to learn user-item and item-item relations in \ac{CML}. Our approach is based on the assumption that \textit{there exists a hierarchical structure} in different relation types and, more specifically, that  \textit{user-item relations are built on top of item-item relations}. The assumption is motivated by the idea that users might predominantly forge their opinion about a new item from this item's relations to others they already interacted with. For instance, some users' opinion on a new French rap song released on a music streaming service could be influenced by their former interactions with French and rap songs.

\subsection{\ac{HLR}}
\label{sec:hlr}

Following the notations of Section \ref{s22}, we continue to denote by $p_u$ and $q_v$ the respective low-dimensional representation vectors of some user $u$ and item $v$. These vectors are of dimension $d$, with $d \ll |\mathcal{U}|$ and $d \ll |\mathcal{V}|$.

\subsubsection{Item-Item Modeling}
\begin{figure*}[t]
  \centering
  \includegraphics[width=0.85\textwidth]{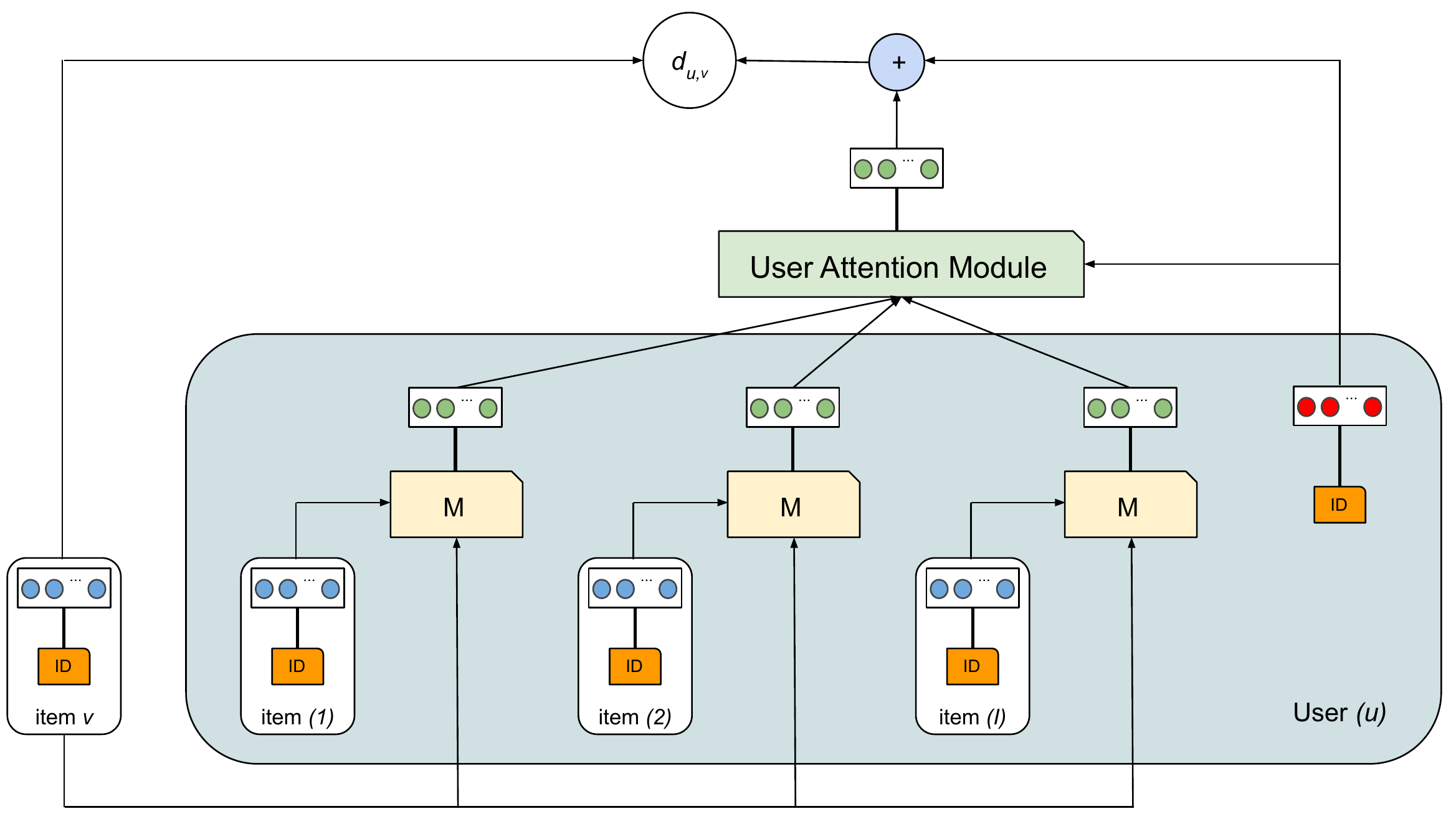}
  \caption{Architecture of \ac{HLR}}
  \label{fig:hlr}
\end{figure*}

Following the above assumption, our proposed \ac{HLR} starts by modeling latent item-item relations. When recommending a candidate item to a user, relations between this item and the user's previous interactions with items will be cornerstones of \ac{HLR}; they are represented as M in Figures \ref{fig:hlr} and \ref{fig:jupiter}.
We model these relations through a \textit{key-value memory network}, as proposed in \cite{tay18:www}. At this stage, we introduce $N$ \textit{keys} vectors $k_i \in \mathbb{R}^d$, stacked up in the matrix $\mathbf{K} \in \mathbb{R}^{N \times d}$, and that \ac{HLR} will learn. Here, $N$ is a hyperparameter representing the number of keys which, in a nutshell, should represent different possible "types of relations". We also introduce the \textit{memory} vectors $m_i \in \mathbb{R}^d$, stacked up in the matrix $\textbf{M} \in \mathbb{R}^{N \times d}$. They are the corresponding \textit{value vectors} of these relation types, as described~thereafter.

\paragraph{Key Addressing $\mathbf{K}$}
First, \ac{HLR} learns which relations exist between two items, as follows:
\begin{enumerate}
    \item Given an item pair $(v_1, v_2)$, a Hadamard product\footnote{We note that other options are also viable to compute joint embedding vectors $s_{v_1,v_2}$, such as some vector averaging or a multi-layer perceptron processing $q_{v_1}$ and $q_{v_2}$. In this work, we use Hadamard products for consistency with \ac{LRML} \cite{tay18:www}, who underlined the empirical effectiveness of~this~simple~approach.} is used to extract the joint embedding between these items:
    \begin{equation}
        s_{v_1,v_2} = q_{v_1} \odot q_{v_2}
    \end{equation}
    where the generated \textit{joint embedding} vector $s_{v_1,v_2} \in \mathbb{R}^d$ is of the same dimension as $q_{v_1}$ and $q_{v_2}$.
    \item Next, we use $s_{v_1,v_2}$ to learn which relation types in $\mathbf{K}$ express the interaction between the two items through an attention vector $w_{v_1,v_2} \in \mathbb{R}^N$, in which each element $w_{i,v_1,v_2} \in \mathbb{R}$ verifies:
    \begin{equation}
        w_{i,v_1,v_2} = s^T_{v_1,v_2}k_i
    \end{equation}
  where we recall that $k_i$ are the $d$-dimensional key vectors from $\mathbf{K}$. Then, attention weights in $w_{v_1,v_2}$ are normalized into a probability distribution, using the \textit{softmax} function. This permits assessing the relative importance of each of the $N$ relations types for a given item pair $(v_1,v_2)$.
\end{enumerate}
\paragraph{Latent Relations via Memories $\mathbf{M}$}
We use the attention vector $w_{v_1,v_2}$ to compute a \textit{weighted} sum of \textit{memory vectors} $m_i \in \mathbb{R}^d$ from the matrix $\mathbf{M} \in \mathbb{R}^{N \times d}$. Each $m_i$ corresponds to the representation vector (to learn) for the relation type $k_i \in \mathbf{K}$. They are building blocks selected to form the \textit{latent relation vector} $r_{v_1, v_2} \in \mathbb{R}^d$ for each item pair $(v_1,v_2)$. Formally, the latent relation vector $r_{v_1, v_2}$ of the $(v_1,v_2)$ pair will be:
\begin{equation}
    r_{v_1, v_2} = \sum^{N}_{i=1} w_{i,v_1,v_2}^{T}m_i.
\end{equation}

\subsubsection{User-Item Modeling}
\label{s322}
Then, we learn latent user-item relations, on top of the item-item ones. More specifically, \ac{HLR} incorporates an \textit{user attention module}, responsible for constructing an adaptive \textit{latent relation vector} $\overline{r}_{u,v} \in \mathbb{R}^d$ between each user $u$ and item $v$, as follows:
\begin{equation}
    \overline{r}_{u,v} = \sum_{j \in \mathcal{N}_u} \alpha(u,v,j) r_{v,j}.
    \label{ruv}
\end{equation}
We recall that $\mathcal{N}_u \subseteq \mathcal{V}$ is the set of items that user $u$ previously interacted with. In equation (\ref{ruv}), $\alpha(u,v,j) \in [0,1]$ denotes the attention weight which aims at capturing the contribution of the item-item relation $r_{v,j}$ to the final user-item latent relation between user $u$ and item $v$. More precisely, we define $\alpha(u,v,j)$ with the standard \textit{softmax} function:
\begin{equation}
    \alpha(u,v,j) = \frac{e^{p_u^T r_{v,j}}}{\sum_{h \in \mathcal{N}_u} e^{p_u^T r_{v,h}}}.
\end{equation}
A large $\alpha(u,v,j)$ indicates a high influence of the relation $r_{v, j}$ in the final representation of the user-item relation $\overline{r}_{u,v}$.
Finally, for recommendation tasks, \textit{user-item similarities} for each user-item pair $(u,v)$ will be evaluated as follows:
\begin{equation}
    d(u, v) = ||p_u + \overline{r}_{u,v} - q_v||^2_2.
\end{equation}

\subsubsection{Optimization}
As other \ac{CML}-based methods \cite{he17:www, tay18:www, park18:icdm}, we minimize a \textit{triplet loss} pulling closer matching user-item pairs in $\mathcal{S}$, and pushing away non-matching pairs up to an $m>0$ distance. Equation \eqref{triplet_loss} is redefined as:
\begin{equation}
    \mathcal{L} = \sum_{(u, v, v^-) \in \mathcal{T}} \Big[||p_u + \overline{r}_{u,v} - q_v||^{2}_2 - ||p_u + \overline{r}_{u,v-} - q_{v^-}||^{2}_2 + m\Big]_+
\end{equation}
where, as in equation \eqref{triplet_loss}, $\mathcal{T}$ denotes the ensemble of \textit{triplets} $(u, v, v^-)$ such as $(u,v) \in \mathcal{S}$ and $(u,v^-) \notin \mathcal{S}$. Intuitively, such a loss penalizes deviations of $p_u + \overline{r}_{u,v}$ from the vector $q_v$. In practice, latent vectors, keys $\mathbf{K}$ and memories $\mathbf{M}$ are learned by iteratively minimizing $\mathcal{L}$, using gradient descent \cite{goodfellow2016deep}, and a triplet sampling scheme described in Section \ref{s43}.

\subsection{\ac{HLR}++}
\label{s33}

\begin{figure*}[t]
  \centering
  \includegraphics[width=0.9\textwidth]{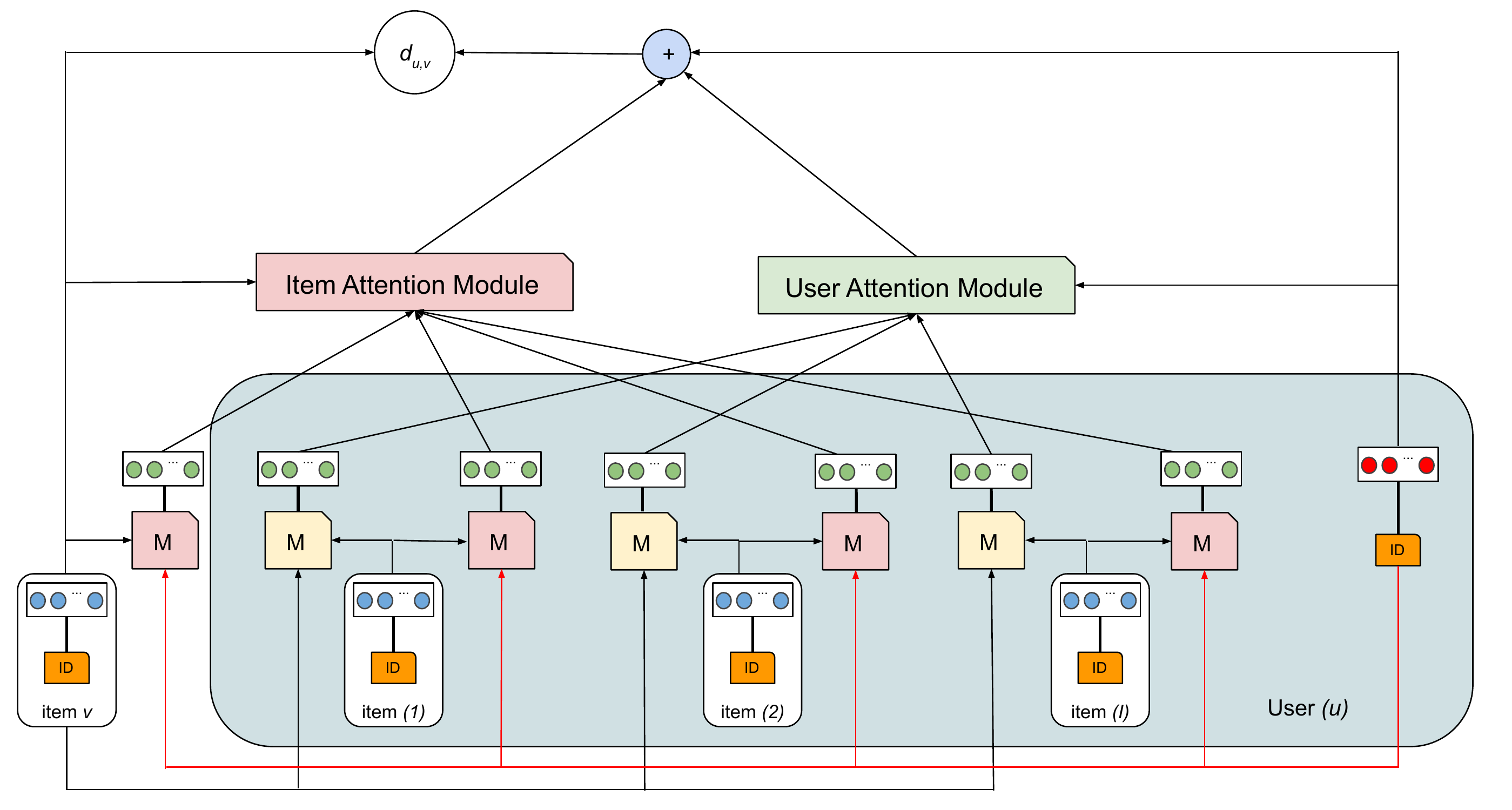}
  \caption{Architecture of \ac{HLR}++}
  \label{fig:jupiter}
\end{figure*}

In our work, we also consider an extension of \ac{HLR}, denoted \ac{HLR}++. As illustrated in Figure \ref{fig:jupiter}, \ac{HLR}++ adds an \textit{item attention module} on top of \ac{HLR}. It acts as a symmetric component w.r.t. the \textit{user attention module} from Section \ref{s322}. This module is responsible for additionally building item-item relations on top of user-item relations. 
It is motivated by the following postulate: some items might be related, not only because of their characteristics, but also because some users consider them together \textit{in a specific context}. For instance, a user might listen to specific songs while running.~In~\ac{HLR}++:
\begin{itemize}
    \item First, we model item-item relations by memory networks, as in \ac{HLR}.
    \item Then, the \textit{item attention module}, together with the pre-existing \textit{user attention module}, symmetrically construct two user-item relation vectors, say $\overline{r}^I_{u,v}$ and $\overline{r}^U_{u,v}$, as in Section \ref{s322} - but averaging over users of a same item~for~$\overline{r}^I_{u,v}$.
    \item Finally, the adaptive relation vector $\overline{r}_{u,v}$ between a user $u$ and a candidate item $v$ is derived by summing up the outputs of both user and item attention modules:
    \begin{equation}
        \overline{r}_{u,v} = \overline{r}^I_{u,v} + \overline{r}^U_{u,v}.
    \end{equation}
\end{itemize}

%%%%%%%%%%%%%%%%%%%%%%%%%%%%%%
%       EXPERIMENTS
%%%%%%%%%%%%%%%%%%%%%%%%%%%%%%
\section{Experimental Evaluation}
\label{s4}

In this section, we empirically evaluate and discuss the performance of our \ac{HLR} and \ac{HLR}++ models on recommendation.
\subsection{Datasets}

For our experimental evaluation, we consider four real-world datasets covering various domains:
\begin{itemize}
    \item \textit{MovieLens} \cite{harper2015movielens}: this dataset gathers user ratings for movies, collected from a movie recommendation service. We binarize explicit rating data on a 5-star scale, keeping the ratings of 4 or higher as positive implicit feedback.
    \item \textit{Echonest} \cite{bertin2011million}: the Echo Nest Taste Profile dataset contains user playcounts for songs from their Million Song Dataset (MSD). The playcount data is binarized by considering values of six streams or higher as implicit feedback.
    \item \textit{Yelp} \cite{yelp2019dataset}: online reviews of businesses on a 5-star scale. Ratings of 4 or 5 are binarized and are~implicit~feedback.  
    \item \textit{Amazon book} \cite{mcauley2015image}: a dataset of consumption records with reviews from \textit{Amazon.com}. We use 5-star ratings from the book category. Ratings are binarized by keeping ratings of 4 and 5 as implicit feedback.
\end{itemize}
We implement a \textit{$k$-core} pre-processing step on each dataset, consisting in a recursive filter until all users and items have at least $k$ interactions, as proposed in \cite{sun20:recsys}. Specifically, we set $k = 10$ for all datasets except for Yelp ($k = 5$). Table \ref{tab:stats} reports some statistics on our four datasets, after this pre-processing step. They are ordered by \textit{density} of the user-item interaction matrix, which is a commonly used factor to assess the difficulty of the recommendation task \cite{silva18:ijcnn}.

\begin{table*}[t]
  \caption{Datasets statistics after pre-processing}
  \label{tab:stats}
  \resizebox{0.7\textwidth}{!}{
  \begin{tabular}{c|ccccc}
    \toprule
    \textbf{Dataset} & \textbf{Number of} & \textbf{Number of} & \textbf{Number of} & \textbf{Density} &\textbf{Median number of} \\
     & \textbf{users} & \textbf{items} & \textbf{ratings} & & \textbf{interactions per user} \\
    \midrule
    \textbf{MovieLens} & 129 757 & 11 508 & 9 911 879 & 0.664\% & 98 \\
    \textbf{Echonest} & 131 495 & 39 874 & 2 476 013 & 0.047\% & 24 \\
    \textbf{Yelp} & 82 166 & 71 949 & 2 077 093 & 0.035\% & 23 \\
    \textbf{Amazon} & 109 730 & 96 421 & 1 405 671 & 0.030\% & 16 \\
    \bottomrule
  \end{tabular}
  }
\end{table*}

%\begin{table}
%  \caption{Statistics of the datasets after applying 10-core filter except Yelp with 5-core}
%  \label{tab:stats}
%  \begin{tabular}{c|ccccc}
%    \toprule
%    \textbf{Dataset} & \textbf{Number of} & \textbf{Number of} & \textbf{Number of} & \textbf{Density} &\textbf{Median number} \\
%     & \textbf{users} & \textbf{items} & \textbf{ratings} & & \textbf{of interactions} \\
%    \midrule
%    MovieLens & 129 757 & 11 508 & 9 911 879 & 0.664\% & 98 \\
%    Echonest & 131 495 & 39 874 & 2 476 013 & 0.047\% & 24 \\
%    Yelp & 82 166 & 71 949 & 2 077 093 & 0.035\% & 23 \\
%    Amazon & 109 730 & 96 421 & 1 405 671 & 0.030\% & 16 \\
%    \bottomrule
%  \end{tabular}
%  \vspace{-4mm}
%\end{table}

\subsection{Evaluation Methodology}

We assess the performance of \ac{HLR} and \ac{HLR}++ on these four datasets, for the top-$K$ recommendation task introduced in Section \ref{s21}, with $K=10$. We closely follow the evaluation protocol proposed by Sun et al. \cite{sun20:recsys}. Specifically, datasets are splitted as follows: 80\% of user interactions are used for training, 10\% for validation and the last 10\% for test. Models must correctly recommend an ordered list of $K=10$ items for which each user should interact with positively, in the validation or test set.
We report five standard evaluation metrics: the Precision and the Recall, for prediction accuracy, as well as the \ac{NDCG}, \ac{MAP} and \ac{MRR} scores as measures of ranking quality. Last, as collaborative filtering is known to be prone to \textit{popularity biases} \cite{steck11:recsys}, consisting in recommending more popular content, we furthermore report the \textit{median popularity} of recommended items. The popularity of an item is defined as the number of users who interacted with it.

\subsection{Models}
\label{s43}

\subsubsection{List of Baselines}
We compare our proposed  \ac{HLR} and \ac{HLR}++ to several baselines.
Foremost, we consider MF-ALS \cite{koren08:sigkdd}, a standard \ac{MF} method modeling the user-item relations using inner products, from the factorization of the training interaction matrix using \textit{alternating least squares} (ALS) \cite{koren2009matrix}. We also consider \ac{CML} \cite{hsieh17:www}, the standard collaborative metric learning approach which does not include any relational translation between user and item vectors. Then, we consider the \ac{LRML} and \ac{TransCF} extensions of \ac{CML}, described in Section \ref{s22}, and capturing heterogeneities in \textit{user-item} interactions. Last, we implement
\ac{AdaCML} \cite{zhang19:dsaa}, described in Section \ref{s23}, which models~\textit{item-item}~relations.

\subsubsection{Implementation Details}
We use Adam optimizer \cite{kingma14:arxiv} for all models. To avoid reconstructing all triplets in $\mathcal{L}$, we adopt a standard \textit{triplet sampling} strategy \cite{hsieh17:www} that approximates losses: specifically, each training instance $(u,v)$ is paired with a single negative sample $(u,v^-)$ randomly picked among all items $v^-$ with which the user $u$ did not interact.

Models are trained for a maximum of 100 epochs. Parameters on the best epoch, i.e. parameters with minimum loss on the validation set, are used for final models. The embedding size $d$ is fixed to 100 while the batch size is set to 1000. Other hyperparameters are tuned on the validation set by grid search: learning rates are amongst $\{0.0002, 0.0005, 0.00075, 0.001\}$; the number of memory slices $N$ is in $\{5, 10, 20, 50\}$; the margin $m$ is in $\{0.2, 0.5, 0.75, 1.0\}$. 

Along with this paper, we publicly release our source code\footnote{\href{https://github.com/deezer/recsys21-hlr}{https://github.com/deezer/recsys21-hlr}} to ensure the reproducibility of our results. Our repository includes an exhaustive table reporting the best hyperparameters for all models. It also reports running times on our machines, showing that \ac{HLR}/\ac{HLR}++ behave on par with the other item-item based model AdaCML.

\subsection{Experimental Results}

Table \ref{performance_results} reports average performances of all models, along with standard deviations over five runs to measure randomness in the training process. Overall, our approach reaches competitive results on three datasets out of four (MovieLens, Echonest and Yelp), while the \ac{AdaCML} baseline ranks first on the Amazon dataset. We thereafter interpret these results.

\begin{table*}[t]
    \caption{Top-10 item recommendation using \ac{HLR}, \ac{HLR}++ and baseline models.}
    \label{performance_results}
    \resizebox{\textwidth}{!}{
    \begin{tabular}{c|c||ccccc||cc}
\toprule
         {\textbf{Datasets}} & {\textbf{Metrics @10   (in \%)}} & {MF-ALS}  & {CML} & {LRML} & {TransCF} & {AdaCML} & {\textbf{\ac{HLR}}} & {\textbf{\ac{HLR}++}} \\
         \midrule
          \midrule
         \multirow{6}{*}{\textbf{MovieLens}} & %{Loc-NDCG} & {60.04 $\pm$ 0.20} & {63.51 $\pm$ 0.19} & {53.45 $\pm$ 0.23} & {64.14 $\pm$ 0.18} & {65.13 $\pm$ 0.19} & {\textbf{65.61 $\pm$ 0.18}} \\ 
         {MAP} & {8.41 $\pm$ 0.09} &  {7.82 $\pm$ 0.08} & {9.40 $\pm$ 0.05} & {2.94 $\pm$ 0.04} & {10.62 $\pm$ 0.08} & {10.97 $\pm$ 0.06} & {\textbf{11.36 $\pm$ 0.06}} \\     
         {} & {MRR} & {55.32 $\pm$ 0.53} & {48.70 $\pm$ 0.39} & {56.54 $\pm$ 0.24} & {22.07 $\pm$ 0.21} & {63.38 $\pm$ 0.24} & {66.01 $\pm$ 0.29} & {\textbf{67.85 $\pm$ 0.24}} \\
         {} & {NDCG} & {20.05 $\pm$ 0.28}  & {18.29 $\pm$ 0.12} & {20.28 $\pm$ 0.06} & {8.36 $\pm$ 0.07} & {23.47 $\pm$ 0.09} & {24.31 $\pm$ 0.13} & {\textbf{25.07 $\pm$ 0.10}} \\
         {} & {Precision} & {15.22 $\pm$ 0.17} & {14.94 $\pm$ 0.17} & {16.37 $\pm$ 0.05} & {7.01 $\pm$ 0.06} & {18.47 $\pm$ 0.09} & {18.80 $\pm$ 0.11} & {\textbf{19.29 $\pm$ 0.07}} \\
         {} & {Recall} & {13.98 $\pm$ 0.29} & {13.22 $\pm$ 0.08} & {13.53 $\pm$ 0.05} & {6.04 $\pm$ 0.05} & {16.96 $\pm$ 0.12} & {17.63 $\pm$ 0.19} & {\textbf{18.26 $\pm$ 0.25}} \\
         \cline{2-9}
         {} & {Popularity} & {4771.6 $\pm$ 56.5} & {1244.8 $\pm$ 45.3} & {1034.6 $\pm$ 18.0} & {1495.5 $\pm$ 28.9} & {835.1 $\pm$ 19.7} & {\textbf{827.8 $\pm$ 20.3}} & {885.4 $\pm$ 7.4} \\
         %{} & {Diversity} & {$\pm$} & {$\pm$} & {$\pm$} & {$\pm$} & {$\pm$} & {$\pm$} \\
         \hline
         \multirow{6}{*}{\textbf{Echonest}} & %{Loc-NDCG} & {57.43 $\pm$ 0.35} & {58.14 $\pm$ 0.66} & {59.03 $\pm$ 0.75} & {58.67 $\pm$ 0.46} & {58.98 $\pm$ 0.29} & {\textbf{61.14 $\pm$ 0.11}} \\ 
         {MAP} & {1.62 $\pm$ 0.03} & {1.97 $\pm$ 0.05} & {1.95 $\pm$ 0.06} & {1.82 $\pm$ 0.07} & {2.72 $\pm$ 0.05} & {3.04 $\pm$ 0.04} & {\textbf{3.28 $\pm$ 0.04}} \\
         {} & {MRR} & {13.44 $\pm$ 0.20}  & {15.82 $\pm$ 0.3} & {15.70 $\pm$ 0.33} & {14.38 $\pm$ 0.52} & {21.25 $\pm$ 0.33} & {23.92 $\pm$ 0.3} & {\textbf{25.70 $\pm$ 0.27}} \\ 
         {} & {NDCG} & {7.12 $\pm$ 0.10} & {8.85 $\pm$ 0.15} & {8.60 $\pm$ 0.13} & {7.90 $\pm$ 0.26} & {11.78 $\pm$ 0.15} & {13.13 $\pm$ 0.16} & {\textbf{13.88 $\pm$ 0.13}} \\
         {} & {Precision} & {3.13 $\pm$ 0.04} & {4.12 $\pm$ 0.07} & {4.02 $\pm$ 0.04} & {3.59 $\pm$ 0.09} & {5.48 $\pm$ 0.05} & {6.18 $\pm$ 0.07} & {\textbf{6.31 $\pm$ 0.04}} \\
         {} & {Recall} & {8.48 $\pm$ 0.11} & {11.29 $\pm$ 0.16} & {10.74 $\pm$ 0.06} & {9.81 $\pm$ 0.27} & {14.89 $\pm$ 0.16} & {16.59 $\pm$ 0.28} & {\textbf{16.98 $\pm$ 0.19}} \\
         \cline{2-9}
         {} & {Popularity} & {511.7 $\pm$ 2.2} & {51.0 $\pm$ 1.4} & {56.4 $\pm$ 0.5} & {55.0 $\pm$ 0.6} & {41.4 $\pm$ 0.8} & {\textbf{33.0 $\pm$ 0.0}} & {35.3 $\pm$ 0.4} \\
         %{} & {Diversity} & {$\pm$} & {$\pm$} & {$\pm$} & {$\pm$} & {$\pm$} & {$\pm$} \\
         \hline
         \multirow{6}{*}{\textbf{Yelp}} & %{Loc-NDCG} & {49.25 $\pm$ 0.49} & {50.16 $\pm$ 0.43} & {$\pm$} & {49.85 $\pm$ 0.47} & {\textbf{51.67 $\pm$ 0.60}} & {51.56 $\pm$ 0.51} \\ 
         {MAP} & {0.44 $\pm$ 0.01} & {0.52 $\pm$ 0.01} & {0.54 $\pm$ 0.02} & {0.15 $\pm$ 0.01} & {0.57 $\pm$ 0.02} & {\textbf{0.63 $\pm$ 0.02}} & {\textbf{0.63 $\pm$ 0.01}} \\
         {} & {MRR} & {4.10 $\pm$ 0.12} & {4.75 $\pm$ 0.1} & {4.99 $\pm$ 0.17} & {1.43 $\pm$ 0.06} & {5.22 $\pm$ 0.17} & {\textbf{5.80 $\pm$ 0.20}} & {\textbf{5.74 $\pm$ 0.11}} \\
         {} & {NDCG} & {2.57 $\pm$ 0.07} & {2.37 $\pm$ 0.08} & {2.45 $\pm$ 0.08} & {0.97 $\pm$ 0.04} & {2.58 $\pm$ 0.04} & {\textbf{2.83 $\pm$ 0.08}} & {\textbf{2.83 $\pm$ 0.04}} \\
         {} & {Precision} & {1.13 $\pm$ 0.02} & {1.49 $\pm$ 0.04} & {1.51 $\pm$ 0.05} & {0.48 $\pm$ 0.02} & {1.60 $\pm$ 0.03} & {\textbf{1.68 $\pm$ 0.02}} & {\textbf{1.67 $\pm$ 0.03}} \\
         {} & {Recall} & {3.64 $\pm$ 0.10} & {2.89 $\pm$ 0.13} & {2.90 $\pm$ 0.11} & {1.53 $\pm$ 0.07} & {3.13 $\pm$ 0.06} & {\textbf{3.35 $\pm$ 0.10}} & {\textbf{3.35 $\pm$ 0.09}} \\
         \cline{2-9}
         {} & {Popularity} & {385.7 $\pm$ 1.1} & {45.2 $\pm$ 0.4} & {46.4 $\pm$ 0.5} & {44.2 $\pm$ 0.4} & {\textbf{40.2 $\pm$ 0.4}} & {68.0 $\pm$ 1.1} & {73.2 $\pm$ 1.3} \\
         %{} & {Diversity} & {$\pm$} & {$\pm$} & {$\pm$} & {$\pm$} & {$\pm$} & {$\pm$} \\
         \hline
         \multirow{6}{*}{\textbf{Amazon}} & %{Loc-NDCG} & {54.13 $\pm$ 0.89} & {54.54 $\pm$ 0.18} & {50.96 $\pm$ 0.6} & {55.39 $\pm$ 0.43} & {54.83 $\pm$ 0.78} & {$\pm$} \\ 
         {MAP} & {0.98 $\pm$ 0.03} & {1.12 $\pm$ 0.05} & {1.06 $\pm$ 0.03} & {0.57 $\pm$ 0.01} & {\textbf{1.33 $\pm$ 0.03}} & {1.11 $\pm$ 0.03} & {1.03 $\pm$ 0.03} \\
         {} & {MRR} & {8.27 $\pm$ 0.20} & {9.45 $\pm$ 0.45} & {9.03 $\pm$ 0.19} & {4.93 $\pm$ 0.09} & {\textbf{11.02 $\pm$ 0.23}} & {9.29 $\pm$ 0.4} & {8.71 $\pm$ 0.26} \\
         {} & {NDCG} & {3.84 $\pm$ 0.11} & {4.72 $\pm$ 0.18} & {4.45 $\pm$ 0.11} & {2.61 $\pm$ 0.06} & {\textbf{5.49 $\pm$ 0.13}} & {4.59 $\pm$ 0.20} & {4.39 $\pm$ 0.14} \\
         {} & {Precision} & {2.10 $\pm$ 0.05} & {2.64 $\pm$ 0.07} & {2.47 $\pm$ 0.06} & {1.48 $\pm$ 0.04} & {\textbf{2.99 $\pm$ 0.05}} & {2.54 $\pm$ 0.08} & {2.46 $\pm$ 0.02} \\
         {} & {Recall} & {4.11 $\pm$ 0.17} & {5.47 $\pm$ 0.17} & {5.09 $\pm$ 0.13} & {3.21 $\pm$ 0.08} & {\textbf{6.32$ \pm$ 0.16}} & {5.26 $\pm$ 0.26} & {5.13 $\pm$ 0.20} \\
         \cline{2-9}
         {} & {Popularity} & {222.7 $\pm$ 2.7} & {33.0 $\pm$ 0.0} & {35.6 $\pm$ 0.5} & {\textbf{26.6 $\pm$ 0.5}} & {31.0 $\pm$ 0.0} & {64.0 $\pm$ 1.8} & {47.6 $\pm$ 0.8} \\
         %{} & {Diversity} & {$\pm$} & {$\pm$} & {$\pm$} & {$\pm$} & {$\pm$} & {$\pm$} \\
         \bottomrule
    \end{tabular}
    }
\end{table*}

\subsubsection{On the Relevance of Relation Modeling} 
We observe that most methods incorporating \textit{relation modeling} (\ac{LRML}, \ac{AdaCML} and our \ac{HLR} and \ac{HLR}++) significantly outperform those who do not (MF-ALS and \ac{CML}) on the densest MovieLens dataset. \ac{AdaCML} and, even more, \ac{HLR} and \ac{HLR}++, remain consistently superior on the sparser Echonest and Yelp datasets. For instance, \ac{HLR}++ reaches top MRR and NDCG scores of 25.70\% and 13.88\% respectively on Echonest, vs. 15.82\% and 8.85\% for the standard \ac{CML}. Our experiments also emphasize how the empirical benefits of relation modeling \textit{tend to decrease on sparser datasets}. Specifically, it is narrowed on Yelp and, on Amazon, relation modeling did not bring any benefit (apart for \ac{AdaCML} - see after). We postulate that, on such datasets with few feedback, modeling user-item or item-item relations by latent memories could actually bring noise and be more~prone~to~overfitting.

\subsubsection{On User-Item Modeling vs Item-Item Modeling} \ac{AdaCML}, modeling only item-item relations, outperforms \ac{LRML} and \ac{TransCF}, modeling only user-item relations, on all four datasets (e.g. with a 10.62\% MAP score on MovieLens, vs. 9.40\% for \ac{LRML} and 2.94\% for  \ac{TransCF} on this same dataset). Learning latent relations between items thus seems to better complement the existing user-item interactions from original data, for this recommendation task. This highlights how user preferences for some items can indeed strongly depend on items they previously interacted with.

\subsubsection{On Hierarchical Latent Relation Modeling} \ac{HLR} and \ac{HLR}++ outperform all baselines on the MovieLens, Echonest and Yelp datasets. This emphasizes the benefits of \textit{jointly} modeling both user-item and item-item latent relations. This also tends to confirm the relevance of our assumption (i.e. that user-item relations are built on top of item-item relations) and therefore of our proposed hierarchical architecture. Nevertheless, we acknowledge that, on the very sparse Amazon dataset, \ac{HLR} and \ac{HLR}++ are outperformed by \ac{AdaCML}. Our explanation is twofold. First of all, modeling user-item relations might be unhelpful - not to say harmful - on this last dataset. This would explain why \ac{LRML} and \ac{TransCF} also fail to surpass a standard \ac{CML}, contrary to \ac{AdaCML} that only focuses on item-item relations. Secondly, \ac{AdaCML} adopts a relatively simple strategy (relying on unweighted averages of inner products) that, despite being surpassed by \ac{HLR} and \ac{HLR}++ on the three other datasets, might be less prone to overfitting in very sparse settings such as Amazon.

\subsubsection{On \ac{HLR}++ vs \ac{HLR}}

The additional \textit{item attention module} of \ac{HLR}++ provides visible gains compared to \ac{HLR} on MovieLens and Echonest (e.g. with a 18.26\% Recall on MovieLens, vs 17.63\% for \ac{HLR}).
This tends to show that items relations could also be built upon user-item ones, for example when considering specific \textit{contexts} (see Section \ref{s33}). Regarding Yelp and Amazon, however, this item attention module does not show any improvement. These datasets might be less prone to contextual interactions, w.r.t. MovieLens (on movies) and Echonest (on music). Indeed, music consumption highly depends on the listening context which may be linked to emotions and mood, time of the day, season, event, location, weather, and user activity \cite{marius12:csr, sergey18:chiir}. In the case of movies, watching context can also be considered an important aspect of consumption \cite{biancalana11:camr}. However, in domains such as business products and books, the influence of these contexts to user behaviors could be lesser, considering that consumption times are higher than for movies and songs (several days to several weeks instead of several minutes or a couple of hours). Such a result is also in line with our previous findings on the limits of complexifying models when dealing with very sparse feedback data.

\subsubsection{On Popularity}
We observe that MF-ALS is more prone to \textit{popularity biases} \cite{steck11:recsys} than \ac{CML} methods on all datasets. \ac{HLR} and \ac{HLR}++ tend to recommend \textit{less popular} content w.r.t. other \ac{CML} extensions on MovieLens and Echonest, which is often viewed as a desirable property \cite{schedl18:jmir,steck11:recsys}. On the contrary, they recommend slightly \textit{more popular} content on Yelp and Amazon; they nonetheless preserve significant reductions in popularity biases w.r.t. standard \ac{MF} methods (e.g. with a 47.6 median popularity on Amazon for \ac{HLR}++, above the 26.6 of \ac{TransCF} but way below the~222.7~of~MF-ALS).

%\subsection{\textcolor{red}{Discussion and Analysis}}
%\textcolor{red}{
%We follow the analysis presented in \cite{tay18:www} to derive qualitative insights of our proposed models by studying how similar between relation vectors are for similar user-item pairs, based on their attributes. Specifically, the latent relation vector $r$ is generated for each user-item pair in the test set. Next, cosine similarity between these relation vectors are computed. The user-item pairs in which the cosine similarity between their relation vectors is the highest are then selected to compute the matching portion of attributes. We investigate the relation vectors in both cases where our proposed models outperform (Movielens 20M) and underperform (Amazon) with respect to baselines, as presented in the table [xxx].
%}

\section{Conclusion}
\label{s5}

In this paper, we introduced a hierarchical \ac{CML} model that, contrary to previous efforts, jointly models latent user-item and item-item relations from implicit data. Our approach leverages memory-based attention networks and builds upon the assumption that user-item relations are built on top of the item-item ones in a hierarchical manner.
We emphasized the empirical effectiveness of our approach at capturing the complex interests of users for recommendation,  outperforming popular alternatives in terms of prediction accuracy and ranking quality on several real-world datasets. On the other hand, we also pointed out that relation modeling does not always improve standard \ac{CML}, especially on sparse datasets with very few interaction data. This leaves the door open for future research on such a challenging setting, where we believe simple approaches might surpass complex models in providing more reliable recommendations.

\bibliographystyle{ACM-Reference-Format}
\bibliography{reference}


\begin{thebibliography}{38}

%%% the Web address.

\ifx \showCODEN    \undefined \def \showCODEN     #1{\unskip}     \fi
\ifx \showDOI      \undefined \def \showDOI       #1{#1}\fi
\ifx \showISBNx    \undefined \def \showISBNx     #1{\unskip}     \fi
\ifx \showISBNxiii \undefined \def \showISBNxiii  #1{\unskip}     \fi
\ifx \showISSN     \undefined \def \showISSN      #1{\unskip}     \fi
\ifx \showLCCN     \undefined \def \showLCCN      #1{\unskip}     \fi
\ifx \shownote     \undefined \def \shownote      #1{#1}          \fi
\ifx \showarticletitle \undefined \def \showarticletitle #1{#1}   \fi
\ifx \showURL      \undefined \def \showURL       {\relax}        \fi
% invisible to TeX
\providecommand\bibfield[2]{#2}
\providecommand\bibinfo[2]{#2}
\providecommand\natexlab[1]{#1}
\providecommand\showeprint[2][]{arXiv:#2}

\bibitem[\protect\citeauthoryear{Bertin-Mahieux, Ellis, Whitman, and
  Lamere}{Bertin-Mahieux et~al\mbox{.}}{2011}]%
        {bertin2011million}
\bibfield{author}{\bibinfo{person}{Thierry Bertin-Mahieux},
  \bibinfo{person}{Daniel~PW Ellis}, \bibinfo{person}{Brian Whitman}, {and}
  \bibinfo{person}{Paul Lamere}.} \bibinfo{year}{2011}\natexlab{}.
\newblock \showarticletitle{The Million Song Dataset}. In
  \bibinfo{booktitle}{\emph{Proceedings of the 12th International Society for
  Music Information Retrieval Conference}}. \bibinfo{pages}{909--916}.
\newblock


\bibitem[\protect\citeauthoryear{Biancalana, Gasparetti, Micarelli, Miola, and
  Sansonetti}{Biancalana et~al\mbox{.}}{2011}]%
        {biancalana11:camr}
\bibfield{author}{\bibinfo{person}{Claudio Biancalana}, \bibinfo{person}{Fabio
  Gasparetti}, \bibinfo{person}{Alessandro Micarelli}, \bibinfo{person}{Alfonso
  Miola}, {and} \bibinfo{person}{Giuseppe Sansonetti}.}
  \bibinfo{year}{2011}\natexlab{}.
\newblock \showarticletitle{Context-aware movie recommendation based on signal
  processing and machine learning}. In \bibinfo{booktitle}{\emph{Proceedings of
  the 2nd Challenge on Context-Aware Movie Recommendation}}.
  \bibinfo{pages}{5--10}.
\newblock


\bibitem[\protect\citeauthoryear{Bobadilla, Ortega, Hernando, and A.}{Bobadilla
  et~al\mbox{.}}{2013}]%
        {bobadilla13:kbs}
\bibfield{author}{\bibinfo{person}{Jesús Bobadilla}, \bibinfo{person}{Fernando
  Ortega}, \bibinfo{person}{Antonio Hernando}, {and}
  \bibinfo{person}{Gutiérrez A.}} \bibinfo{year}{2013}\natexlab{}.
\newblock \showarticletitle{Recommender Systems Survey}.
\newblock \bibinfo{journal}{\emph{Know.-Based Sys.}}  \bibinfo{volume}{46}
  (\bibinfo{year}{2013}), \bibinfo{pages}{109--132}.
\newblock


\bibitem[\protect\citeauthoryear{Bordes, Usunier, Garcia-Duran, Weston, and
  Yakhnenko}{Bordes et~al\mbox{.}}{2013}]%
        {bordes13:nips}
\bibfield{author}{\bibinfo{person}{Antoine Bordes}, \bibinfo{person}{Nicolas
  Usunier}, \bibinfo{person}{Alberto Garcia-Duran}, \bibinfo{person}{Jason
  Weston}, {and} \bibinfo{person}{Oksana Yakhnenko}.}
  \bibinfo{year}{2013}\natexlab{}.
\newblock \showarticletitle{Translating Embeddings for Modeling
  Multi-Relational Data}. In \bibinfo{booktitle}{\emph{Advances in Neural
  Information Processing Systems}}. \bibinfo{pages}{1--9}.
\newblock


\bibitem[\protect\citeauthoryear{Briand, Salha-Galvan, Bendada, Morlon, and
  Tran}{Briand et~al\mbox{.}}{2021}]%
        {briand21:kdd}
\bibfield{author}{\bibinfo{person}{L{\'e}a Briand}, \bibinfo{person}{Guillaume
  Salha-Galvan}, \bibinfo{person}{Walid Bendada}, \bibinfo{person}{Mathieu
  Morlon}, {and} \bibinfo{person}{Viet-Anh Tran}.}
  \bibinfo{year}{2021}\natexlab{}.
\newblock \showarticletitle{A Semi-Personalized System for User Cold Start
  Recommendation on Music Streaming Apps}.
\newblock \bibinfo{journal}{\emph{Proceedings of the 27th ACM SIGKDD Conference
  on Knowledge Discovery and Data Mining}}.
\newblock


\bibitem[\protect\citeauthoryear{Chen, Moore, Turnbull, and Joachims}{Chen
  et~al\mbox{.}}{2012}]%
        {chen12:sigkdd}
\bibfield{author}{\bibinfo{person}{Shuo Chen}, \bibinfo{person}{Josh~L. Moore},
  \bibinfo{person}{Douglas Turnbull}, {and} \bibinfo{person}{Thorsten
  Joachims}.} \bibinfo{year}{2012}\natexlab{}.
\newblock \showarticletitle{Playlist Prediction via Metric Embedding}. In
  \bibinfo{booktitle}{\emph{Proceedings of the 18th ACM SIGKDD International
  Conference on Knowledge Discovery and Data Mining}}.
  \bibinfo{pages}{714--722}.
\newblock


\bibitem[\protect\citeauthoryear{Feng, Li, Zeng, Cong, Chee, and Yuan}{Feng
  et~al\mbox{.}}{2015}]%
        {feng15:ijcai}
\bibfield{author}{\bibinfo{person}{Shanshan Feng}, \bibinfo{person}{Xutao Li},
  \bibinfo{person}{Yifeng Zeng}, \bibinfo{person}{Gao Cong},
  \bibinfo{person}{Yeow~Meng Chee}, {and} \bibinfo{person}{Quan Yuan}.}
  \bibinfo{year}{2015}\natexlab{}.
\newblock \showarticletitle{Personalized Ranking Metric Embedding for Next New
  POI Recommendation}. In \bibinfo{booktitle}{\emph{Proceedings of the 24th
  International Conference on Artificial Intelligence}}.
  \bibinfo{pages}{2069--2075}.
\newblock


\bibitem[\protect\citeauthoryear{Goodfellow, Bengio, and Courville}{Goodfellow
  et~al\mbox{.}}{2016}]%
        {goodfellow2016deep}
\bibfield{author}{\bibinfo{person}{I. Goodfellow}, \bibinfo{person}{Y. Bengio},
  {and} \bibinfo{person}{A. Courville}.} \bibinfo{year}{2016}\natexlab{}.
\newblock \bibinfo{booktitle}{\emph{Deep Learning}}.
\newblock \bibinfo{publisher}{MIT Press}.
\newblock


\bibitem[\protect\citeauthoryear{Harper and Konstan}{Harper and
  Konstan}{2015}]%
        {harper2015movielens}
\bibfield{author}{\bibinfo{person}{F~Maxwell Harper} {and}
  \bibinfo{person}{Joseph~A Konstan}.} \bibinfo{year}{2015}\natexlab{}.
\newblock \showarticletitle{The MovieLens Datasets: History and Context}.
\newblock \bibinfo{journal}{\emph{ACM Trans. on Int. Int. Sys.}}
  \bibinfo{volume}{5}, \bibinfo{number}{4} (\bibinfo{year}{2015}),
  \bibinfo{pages}{1--19}.
\newblock


\bibitem[\protect\citeauthoryear{He, Lizi~Liao, Nie, Hu, and Chua}{He
  et~al\mbox{.}}{2017}]%
        {he17:www}
\bibfield{author}{\bibinfo{person}{Xiangnan He}, \bibinfo{person}{Hanwang~Zhang
  Lizi~Liao}, \bibinfo{person}{Liqiang Nie}, \bibinfo{person}{Xia Hu}, {and}
  \bibinfo{person}{Tat-Seng Chua}.} \bibinfo{year}{2017}\natexlab{}.
\newblock \showarticletitle{Neural Collaborative Filtering}. In
  \bibinfo{booktitle}{\emph{In Proceedings of the 26th International Conference
  on World Wide Web}}. \bibinfo{pages}{173--182}.
\newblock


\bibitem[\protect\citeauthoryear{Hsieh, Yang, Cui, Lin, Belongie, and
  Estrin}{Hsieh et~al\mbox{.}}{2017}]%
        {hsieh17:www}
\bibfield{author}{\bibinfo{person}{Cheng-Kang Hsieh}, \bibinfo{person}{Longqi
  Yang}, \bibinfo{person}{Yin Cui}, \bibinfo{person}{Tsung-Yi Lin},
  \bibinfo{person}{Serge~J. Belongie}, {and} \bibinfo{person}{Deborah Estrin}.}
  \bibinfo{year}{2017}\natexlab{}.
\newblock \showarticletitle{Collaborative Metric Learning}. In
  \bibinfo{booktitle}{\emph{Proceedings of the World Wide Web Conference}}.
  \bibinfo{pages}{193--201}.
\newblock


\bibitem[\protect\citeauthoryear{Hu, Koren, and Volinsky}{Hu
  et~al\mbox{.}}{2008}]%
        {hu08:icdm}
\bibfield{author}{\bibinfo{person}{Yifan Hu}, \bibinfo{person}{Yehuda Koren},
  {and} \bibinfo{person}{Chris Volinsky}.} \bibinfo{year}{2008}\natexlab{}.
\newblock \showarticletitle{Collaborative Filtering for Implicit Feedback
  Datasets}. In \bibinfo{booktitle}{\emph{Proceedings of the 8th IEEE
  International Conference on Data Mining}}. \bibinfo{pages}{263--272}.
\newblock


\bibitem[\protect\citeauthoryear{Kabbur, Ning, and Karypis}{Kabbur
  et~al\mbox{.}}{2013}]%
        {santosh13:sigkdd}
\bibfield{author}{\bibinfo{person}{Santosh Kabbur}, \bibinfo{person}{Xia Ning},
  {and} \bibinfo{person}{George Karypis}.} \bibinfo{year}{2013}\natexlab{}.
\newblock \showarticletitle{Fism: Factored Item Similarity Models for Top-N
  Recommender Systems}. In \bibinfo{booktitle}{\emph{Proceedings of the 19th
  ACM SIGKDD International Conference on Knowledge Discovery and Data Mining}}.
  \bibinfo{pages}{659--667}.
\newblock


\bibitem[\protect\citeauthoryear{Khoshneshin and Street}{Khoshneshin and
  Street}{2010}]%
        {khoshneshin10:ijcai}
\bibfield{author}{\bibinfo{person}{Mohammad Khoshneshin} {and}
  \bibinfo{person}{Nick~W. Street}.} \bibinfo{year}{2010}\natexlab{}.
\newblock \showarticletitle{Collaborative Filtering via Euclidean Embedding}.
  In \bibinfo{booktitle}{\emph{Proceedings of the ACM Conference on Recommender
  Systems}}. \bibinfo{pages}{87--94}.
\newblock


\bibitem[\protect\citeauthoryear{Kingma and Ba}{Kingma and Ba}{2015}]%
        {kingma14:arxiv}
\bibfield{author}{\bibinfo{person}{Diederik~P Kingma} {and}
  \bibinfo{person}{Jimmy Ba}.} \bibinfo{year}{2015}\natexlab{}.
\newblock \showarticletitle{Adam: A Method for Stochastic Optimization}. In
  \bibinfo{booktitle}{\emph{Proc. of the 3rd Int. Conference on Learning
  Representations}}.
\newblock


\bibitem[\protect\citeauthoryear{Koren}{Koren}{2008}]%
        {koren08:sigkdd}
\bibfield{author}{\bibinfo{person}{Yehuda Koren}.}
  \bibinfo{year}{2008}\natexlab{}.
\newblock \showarticletitle{Factorization Meets the Neighborhood: a
  Multifaceted Collaborative Filtering Model}. In
  \bibinfo{booktitle}{\emph{Proceedings of the 14th ACM SIGKDD International
  Conference on Knowledge Discovery and Data Mining}}.
  \bibinfo{pages}{426--434}.
\newblock


\bibitem[\protect\citeauthoryear{Koren, Bell, and Volinsky}{Koren
  et~al\mbox{.}}{2009}]%
        {koren2009matrix}
\bibfield{author}{\bibinfo{person}{Yehuda Koren}, \bibinfo{person}{Robert
  Bell}, {and} \bibinfo{person}{Chris Volinsky}.}
  \bibinfo{year}{2009}\natexlab{}.
\newblock \showarticletitle{Matrix Factorization Techniques for Recommender
  Systems}.
\newblock \bibinfo{journal}{\emph{Computer}} \bibinfo{volume}{42},
  \bibinfo{number}{8} (\bibinfo{year}{2009}), \bibinfo{pages}{30--37}.
\newblock


\bibitem[\protect\citeauthoryear{Marius and Ricci}{Marius and Ricci}{2012}]%
        {marius12:csr}
\bibfield{author}{\bibinfo{person}{Kaminskas Marius} {and}
  \bibinfo{person}{Francesco Ricci}.} \bibinfo{year}{2012}\natexlab{}.
\newblock \showarticletitle{Contextual music information retrieval and
  recommendation: State of the art and challenges}.
\newblock \bibinfo{journal}{\emph{Computer Science Review}}
  \bibinfo{volume}{6}, \bibinfo{number}{2-3} (\bibinfo{year}{2012}),
  \bibinfo{pages}{89--119}.
\newblock


\bibitem[\protect\citeauthoryear{McAuley, Targett, Shi, and Van
  Den~Hengel}{McAuley et~al\mbox{.}}{2015}]%
        {mcauley2015image}
\bibfield{author}{\bibinfo{person}{Julian McAuley},
  \bibinfo{person}{Christopher Targett}, \bibinfo{person}{Qinfeng Shi}, {and}
  \bibinfo{person}{Anton Van Den~Hengel}.} \bibinfo{year}{2015}\natexlab{}.
\newblock \showarticletitle{Image-Based Recommendations on Styles and
  Substitutes}. In \bibinfo{booktitle}{\emph{Proceedings of the 38th ACM SIGIR
  Conference on Research and Development in Information Retrieval}}.
  \bibinfo{pages}{43--52}.
\newblock


\bibitem[\protect\citeauthoryear{Park, Kim, Xie, and Yu}{Park
  et~al\mbox{.}}{2018}]%
        {park18:icdm}
\bibfield{author}{\bibinfo{person}{Chanyoung Park}, \bibinfo{person}{Donghyun
  Kim}, \bibinfo{person}{Xing Xie}, {and} \bibinfo{person}{Hwanjo Yu}.}
  \bibinfo{year}{2018}\natexlab{}.
\newblock \showarticletitle{Collaborative Translational Metric Learning}. In
  \bibinfo{booktitle}{\emph{Proceedings of the 2018 IEEE International
  Conference on Data Mining (ICDM)}}.
\newblock


\bibitem[\protect\citeauthoryear{Paudel, Luck, and Bernstein}{Paudel
  et~al\mbox{.}}{2019}]%
        {paudel18:wsdm}
\bibfield{author}{\bibinfo{person}{Bibek Paudel}, \bibinfo{person}{Sandro
  Luck}, {and} \bibinfo{person}{Abraham Bernstein}.}
  \bibinfo{year}{2019}\natexlab{}.
\newblock \showarticletitle{Loss Aversion in Recommender Systems: Utilizing
  Negative User Preference to Improve Recommendation Quality}. In
  \bibinfo{booktitle}{\emph{Proceedings of the 12th ACM International
  Conference on Web Search and Data Mining}}.
\newblock


\bibitem[\protect\citeauthoryear{Rendle, Freudenthaler, Gantner, and
  Schmidt-Thieme}{Rendle et~al\mbox{.}}{2009}]%
        {rendle09:auai}
\bibfield{author}{\bibinfo{person}{Steffen Rendle}, \bibinfo{person}{Christoph
  Freudenthaler}, \bibinfo{person}{Zeno Gantner}, {and} \bibinfo{person}{Lars
  Schmidt-Thieme}.} \bibinfo{year}{2009}\natexlab{}.
\newblock \showarticletitle{BPR: Bayesian Personalized Ranking from Implicit
  Feedback}. In \bibinfo{booktitle}{\emph{Proceedings of the 25th Conference on
  Uncertainty in Artificial Intelligence}}.
\newblock


\bibitem[\protect\citeauthoryear{Schafer, Konstan, and Riedl}{Schafer
  et~al\mbox{.}}{2001}]%
        {schafer01:dmkd}
\bibfield{author}{\bibinfo{person}{J~Ben Schafer}, \bibinfo{person}{Joseph~A
  Konstan}, {and} \bibinfo{person}{John Riedl}.}
  \bibinfo{year}{2001}\natexlab{}.
\newblock \showarticletitle{E-commerce Recommendation Applications}.
\newblock \bibinfo{journal}{\emph{Data Mining and Knowledge Discovery}}
  \bibinfo{volume}{5}, \bibinfo{number}{1} (\bibinfo{year}{2001}),
  \bibinfo{pages}{115--153}.
\newblock


\bibitem[\protect\citeauthoryear{Schedl, Zamani, Chen, Deldjoo, and
  Elahi}{Schedl et~al\mbox{.}}{2018}]%
        {schedl18:jmir}
\bibfield{author}{\bibinfo{person}{Markus Schedl}, \bibinfo{person}{Hamed
  Zamani}, \bibinfo{person}{Ching-Wei Chen}, \bibinfo{person}{Yashar Deldjoo},
  {and} \bibinfo{person}{Mehdi Elahi}.} \bibinfo{year}{2018}\natexlab{}.
\newblock \showarticletitle{Current Challenges and Visions in Music Recommender
  Systems Research}.
\newblock \bibinfo{journal}{\emph{International Journal of Multimedia
  Information Retrieval}} \bibinfo{volume}{7}, \bibinfo{number}{2}
  (\bibinfo{year}{2018}), \bibinfo{pages}{95--116}.
\newblock


\bibitem[\protect\citeauthoryear{Sergey and Agichtein}{Sergey and
  Agichtein}{2018}]%
        {sergey18:chiir}
\bibfield{author}{\bibinfo{person}{Volokhin Sergey} {and}
  \bibinfo{person}{Eugene Agichtein}.} \bibinfo{year}{2018}\natexlab{}.
\newblock \showarticletitle{Understanding music listening intents during daily
  activities with implications for contextual music recommendation}. In
  \bibinfo{booktitle}{\emph{Proceedings of the 2018 Conference on Human
  Information Interaction \& Retrieval}}. \bibinfo{pages}{313--316}.
\newblock


\bibitem[\protect\citeauthoryear{Silva, Junior, and Caloba}{Silva
  et~al\mbox{.}}{2018}]%
        {silva18:ijcnn}
\bibfield{author}{\bibinfo{person}{Joao~Felipe Silva},
  \bibinfo{person}{Natanael~Moura Junior}, {and} \bibinfo{person}{Luiz
  Caloba}.} \bibinfo{year}{2018}\natexlab{}.
\newblock \showarticletitle{Effects of Data Sparsity on Recommender Systems
  based on Collaborative Filtering}. In \bibinfo{booktitle}{\emph{Proceedings
  of the 2018 International Joint Conference on Neural Networks}}.
  \bibinfo{pages}{1--8}.
\newblock


\bibitem[\protect\citeauthoryear{Song, Nie, Han, and Li}{Song
  et~al\mbox{.}}{2017}]%
        {song17:aaai}
\bibfield{author}{\bibinfo{person}{Kun Song}, \bibinfo{person}{Feiping Nie},
  \bibinfo{person}{Junwei Han}, {and} \bibinfo{person}{Xuelong Li}.}
  \bibinfo{year}{2017}\natexlab{}.
\newblock \showarticletitle{Parameter Free Large Margin Nearest Neighbor for
  Distance Metric Learning}. In \bibinfo{booktitle}{\emph{Proceedings of the
  31st AAAI Conference on Artificial Intelligence}}.
\newblock


\bibitem[\protect\citeauthoryear{Steck}{Steck}{2011}]%
        {steck11:recsys}
\bibfield{author}{\bibinfo{person}{Harald Steck}.}
  \bibinfo{year}{2011}\natexlab{}.
\newblock \showarticletitle{Item Popularity and Recommendation Accuracy}. In
  \bibinfo{booktitle}{\emph{Proceedings of the Fifth ACM Conference on
  Recommender systems}}. \bibinfo{pages}{125--132}.
\newblock


\bibitem[\protect\citeauthoryear{Sun, Yu, Fang, Yang, Qu, Zhang, and Geng}{Sun
  et~al\mbox{.}}{2020}]%
        {sun20:recsys}
\bibfield{author}{\bibinfo{person}{Zhu Sun}, \bibinfo{person}{Di Yu},
  \bibinfo{person}{Hui Fang}, \bibinfo{person}{Jie Yang},
  \bibinfo{person}{Xinghua Qu}, \bibinfo{person}{Jie Zhang}, {and}
  \bibinfo{person}{Cong Geng}.} \bibinfo{year}{2020}\natexlab{}.
\newblock \showarticletitle{Are We Evaluating Rigorously? Benchmarking
  Recommendation for Reproducible Evaluation and Fair Comparison}. In
  \bibinfo{booktitle}{\emph{Proceedings of the 14th ACM Conference on
  Recommender Systems}}. \bibinfo{pages}{23--32}.
\newblock


\bibitem[\protect\citeauthoryear{Tay, Luu, and Siu}{Tay et~al\mbox{.}}{2018}]%
        {tay18:www}
\bibfield{author}{\bibinfo{person}{Yi Tay}, \bibinfo{person}{Anh~Tuan Luu},
  {and} \bibinfo{person}{Cheung~Hui Siu}.} \bibinfo{year}{2018}\natexlab{}.
\newblock \showarticletitle{Latent Relational Metric Learning via Memory-Based
  Attention for Collaborative Ranking}. In
  \bibinfo{booktitle}{\emph{Proceedings of the 2018 World Wide Web
  Conference}}. \bibinfo{pages}{729--739}.
\newblock


\bibitem[\protect\citeauthoryear{Tran, Tay, Zhang, Cong, and Li}{Tran
  et~al\mbox{.}}{2020}]%
        {tran20:wsdm}
\bibfield{author}{\bibinfo{person}{Lucas~Vinh Tran}, \bibinfo{person}{Yi Tay},
  \bibinfo{person}{Shuai Zhang}, \bibinfo{person}{Gao Cong}, {and}
  \bibinfo{person}{Xiaoli Li}.} \bibinfo{year}{2020}\natexlab{}.
\newblock \showarticletitle{A Boosting Metric Learning Approach in Hyperbolic
  Space for Recommender Systems}. In \bibinfo{booktitle}{\emph{Proceedings of
  the 13th International Conference on Web Search and Data Mining}}.
  \bibinfo{pages}{609--617}.
\newblock


\bibitem[\protect\citeauthoryear{Tran, Hennequin, Royo-Letelier, and
  Moussallam}{Tran et~al\mbox{.}}{2019}]%
        {tran19:sigir}
\bibfield{author}{\bibinfo{person}{Viet-Anh Tran}, \bibinfo{person}{Romain
  Hennequin}, \bibinfo{person}{Jimena Royo-Letelier}, {and}
  \bibinfo{person}{Manuel Moussallam}.} \bibinfo{year}{2019}\natexlab{}.
\newblock \showarticletitle{Improving Collaborative Metric Learning with
  Efficient Negative Sampling}. In \bibinfo{booktitle}{\emph{Proceedings of the
  42nd International ACM SIGIR Conference on Research and Development in
  Information Retrieval}}. \bibinfo{pages}{1201--1204}.
\newblock


\bibitem[\protect\citeauthoryear{Weinberger and Saul}{Weinberger and
  Saul}{2009}]%
        {weinberger:jmlr09}
\bibfield{author}{\bibinfo{person}{Kilian~Q. Weinberger} {and}
  \bibinfo{person}{Lawrence~K. Saul}.} \bibinfo{year}{2009}\natexlab{}.
\newblock \showarticletitle{Distance Metric Learning for Large Margin Nearest
  Neighbor Classification}.
\newblock \bibinfo{journal}{\emph{Journal of Machine Learning Research}}
  \bibinfo{volume}{10} (\bibinfo{year}{2009}), \bibinfo{pages}{207--244}.
\newblock


\bibitem[\protect\citeauthoryear{Yelp}{Yelp}{2014}]%
        {yelp2019dataset}
\bibfield{author}{\bibinfo{person}{Yelp}.} \bibinfo{year}{2014}\natexlab{}.
\newblock \showarticletitle{Dataset available on:
  https://www.yelp.com/dataset/}.
\newblock  (\bibinfo{year}{2014}).
\newblock


\bibitem[\protect\citeauthoryear{Zhang, Yao, Sun, and Tay}{Zhang
  et~al\mbox{.}}{2019a}]%
        {zhang2019deep}
\bibfield{author}{\bibinfo{person}{Shuai Zhang}, \bibinfo{person}{Lina Yao},
  \bibinfo{person}{Aixin Sun}, {and} \bibinfo{person}{Yi Tay}.}
  \bibinfo{year}{2019}\natexlab{a}.
\newblock \showarticletitle{Deep Learning based Recommender System: A Survey
  and New Perspectives}.
\newblock \bibinfo{journal}{\emph{Comput. Surveys}} \bibinfo{volume}{52},
  \bibinfo{number}{1} (\bibinfo{year}{2019}), \bibinfo{pages}{1--38}.
\newblock


\bibitem[\protect\citeauthoryear{Zhang, Yao, Tay, Xu, Zhang, and Zhu}{Zhang
  et~al\mbox{.}}{2018}]%
        {zhang18:corr}
\bibfield{author}{\bibinfo{person}{Shuai Zhang}, \bibinfo{person}{Lina Yao},
  \bibinfo{person}{Yi Tay}, \bibinfo{person}{Xiwei Xu}, \bibinfo{person}{Xiang
  Zhang}, {and} \bibinfo{person}{Liming Zhu}.} \bibinfo{year}{2018}\natexlab{}.
\newblock \showarticletitle{Metric Factorization: Recommendation beyond Matrix
  Factorization}.
\newblock \bibinfo{journal}{\emph{arXiv preprint arXiv:1802.04606}}.
\newblock


\bibitem[\protect\citeauthoryear{Zhang, Zhao, Liu, Xu, Fang, Zhao, Sheng, and
  Cui}{Zhang et~al\mbox{.}}{2019b}]%
        {zhang19:dsaa}
\bibfield{author}{\bibinfo{person}{Tingting Zhang}, \bibinfo{person}{Pengpeng
  Zhao}, \bibinfo{person}{Yanchi Liu}, \bibinfo{person}{Jiajie Xu},
  \bibinfo{person}{Junhua Fang}, \bibinfo{person}{Lei Zhao},
  \bibinfo{person}{Victor~S. Sheng}, {and} \bibinfo{person}{Zhiming Cui}.}
  \bibinfo{year}{2019}\natexlab{b}.
\newblock \showarticletitle{AdaCML: Adaptive Collaborative Metric Learning for
  Recommendation}. In \bibinfo{booktitle}{\emph{Proc. of the International
  Conference on Database Systems for Advanced Applications}}.
  \bibinfo{pages}{301--316}.
\newblock


\bibitem[\protect\citeauthoryear{Zhou, Liu, Lian, and Xie}{Zhou
  et~al\mbox{.}}{2019}]%
        {zhou19:ijcai}
\bibfield{author}{\bibinfo{person}{Xiao Zhou}, \bibinfo{person}{Danyang Liu},
  \bibinfo{person}{Jianxun Lian}, {and} \bibinfo{person}{Xing Xie}.}
  \bibinfo{year}{2019}\natexlab{}.
\newblock \showarticletitle{Collaborative Metric Learning with Memory Network
  for Multi-Relational Recommender Systems}. In
  \bibinfo{booktitle}{\emph{Proceedings of the 28th International Joint
  Conference on Artificial Intelligence}}. \bibinfo{pages}{4454--4460}.
\newblock


\end{thebibliography}

\end{document}